\documentclass[conference,a4paper,10pt]{IEEEtran} %,compsoc
\usepackage{times,subfigure}

\usepackage{graphicx,multirow}
\usepackage{latexsym,color,float,amsmath,amssymb}
\IEEEoverridecommandlockouts

\begin{document}

\title{Wireless Video Transmission with Over-the-Air Packet Mixing\thanks{Antonios Argyriou would like to acknowledge the support from the European Commission through the Marie Curie Intra-European Fellowship WINIE-273041 and the STREP project CONECT
(FP7ICT257616).}}%

\author{\IEEEauthorblockN{Antonios Argyriou}
\IEEEauthorblockA{Department of Computer and Communication Engineering,
 University of Thessaly, Volos, 38221, Greece%\\ %Email: anargyr@ieee.org
 }
 }

%\IEEEpubid{0000--0000/00\$00.00~\copyright~200X IEEE}

\maketitle%

%\graphicspath{{figures/}}

\markboth{Submitted to IEEE Transactions on Communications, \today}{Compressed Video Transmission in Cooperatively Interfering Distributed Wireless Networks}

\begin{abstract}
In this paper, we propose a system for wireless video transmission with a wireless physical layer (PHY) that supports cooperative forwarding of interfered/superimposed packets. Our system model considers multiple and independent unicast transmissions between network nodes while a number of them serve as relays of the interfered/superimposed signals. For this new PHY the average transmission rate that each node can achieve is estimated first. Next, we formulate a utility optimization framework for the video transmission problem and we show that it can be simplified due to the features of the new PHY. Simulation results reveal the system operating regions for which superimposing wireless packets is a better choice than a typical cooperative PHY.
\end{abstract}


%\tableofcontents
%\newpage

\section{Introduction}
Even though a plethora of problems associated with wireless video transmission have been well-investigated by the research community~\cite{mihaela-book}, the proliferation of wireless high-quality video-capturing devices (e.g. smartphones) set even more demanding requirements to the communication layer. Contrary to elastic video distribution mechanisms (e.g. video download) that have seen huge benefits from the increases in network capacity, streaming in wireless communication systems still suffers from several problems: Channel errors are difficult to be corrected in real-time in a way that the impairment is not observable by the user~\cite{mihaela-book}. The classic mechanisms of forward error correction (FEC) or automatic repeat request (ARQ) introduce significant overheads that cannot address entirely the problem of error resiliency due to the dynamic nature of the wireless channels. Furthermore, asymmetry of multi-hop wireless networks exacerbates the problem of video transmission through such a path~\cite{argyriou:tmm-2009}. Bandwidth fluctuations in the wireless case create very frequently the well known video stuttering problems requiring thus bigger playback buffers and larger playback delay. One avenue for significantly minimizing the impact of the aforementioned issues, is to exploit the presence of the higher number of wireless devices that are in close proximity. This can happen if cooperation between the nodes is employed to the advantage of all of them.

%\begin{figure}[t]
%\begin{center}
%%   \subfigure[Many-to-Many]{
%%\subfigure[]{
%\includegraphics[keepaspectratio,width = 0.35\linewidth]{ancol-example-topo.eps}
%%\subfigure[]{
%\includegraphics[keepaspectratio,width = 0.6\linewidth]{ancol-example-time.eps}
%  \caption{A simple example of cooperative interfering transmissions that exercises the proposed PHY.}
%  \label{fig:ancol-example}
%\end{center}
%\end{figure}

Wireless cooperative transmission of video streams has been studied extensively the last few years. We review representative important works next. One interesting approach is to exploit the structure of layered encoded video in order to transmit a subset of the stream depending on the changing network conditions~\cite{alay08}. Video multicast with a more advanced cooperative scheme was considered in~\cite{alay10}. In that work network nodes that are willing to cooperate (relays) employ distributed space-time codes in order to reduce the bit error rate (BER) at physical layer (PHY) when multiple receivers are involved. More recently, cooperation through wireless packet-based network coding was also studied as scheme for improving video transmission. The case of algebraic network coding and video transmission was studied in~\cite{markopoulou-netcod-video}. In that work the authors employ linear network codes for mixing video packets before transmission to a group of senders. In~\cite{liu-cheung09} the authors combine network coding and cooperation and present a rate-distortion optimized and network-coding-based, cooperative peer-to-peer packet repair solution for the multi-stream WWAN video broadcast. Also in~\cite{chou-coding} the authors studied the use of broadcasting from multiple stations for wireless video transmission but without a systematic cooperative protocol for allowing and exploiting interfering transmissions.

The approach we investigate, even though it uses node cooperation, is fundamentally different from the aforementioned works since it exploits wireless packet mixing (superimposed packets or interfering transmissions) for significantly increasing the throughput of the video transmitting users. Physical layer network coding (PLNC) is another term frequently used for this technique. Consider a simple bidirectional traffic scenario where two nodes desire to transmit a packet to each other through the help of an intermediate relay. If PLNC is employed the task of the relay is to forward the mixed signal to the two destinations. Then, both destinations can decode at the PHY the mixed signal since they already have the information packet they transmitted themselves~\cite{dankberg97,zhang:physical-layer-nc,katabi07a}. By overhearing signals and with the help of a relay, more transmissions can occur per time unit increasing thus throughput. The need for a-priori packet knowledge at the receivers was removed in~\cite{argyriou:twc-ancol} where joint decoding of the relayed signals was employed for this purpose. However, for the particular class of packet-based video communication systems the idea of allowing the packets to interfere might not be always beneficial. The reason is that with video communications when a certain packet is transmitted the importance of this packet might be completely different from subsequent or previous packets. Therefore, in the case of video payload the problem that has to be solved is to identify the conditions for allowing specific packet transmissions to interfere given a certain rate-distortion (RD) characterization of the transmitted bitstream and playback delay requirement of the streaming application.

\begin{figure}[t]
\begin{center}
%   \subfigure[Many-to-Many]{
%\subfigure[ ]{
\includegraphics[keepaspectratio,width = 0.8\linewidth]{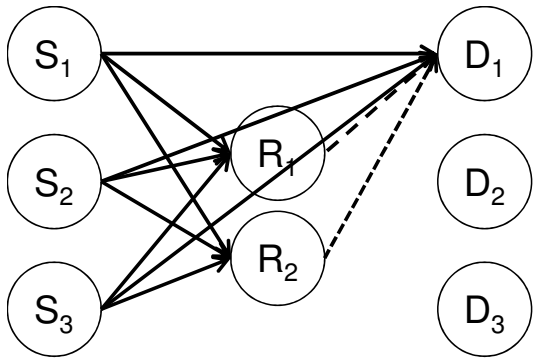}
%}
%\subfigure[Cooperative multicast video transmission with the help of relay nodes $R_1$ and $R_2$. Lines are not shown for $D_2$ to avoid clogging the figure. ]{   \includegraphics[keepaspectratio,width = 0.8\linewidth]{ancol-extension-topo-v2.eps}}
  \caption{The basic network scenario considered in this paper is cooperative unicast video transmission with the help of relay nodes $R_1$ and $R_2$. Different line styles indicate transmission in a different time slot.}
  \label{fig:ancol-video-topo}
\end{center}
\end{figure}

\section{System Model and Overview}
\label{section:system-model}
In this paper, we study a wireless ad hoc network where all nodes can send data to each other and also be potential relays. Multiple unicast flows are assumed. The proposed cooperative PHY protocol optimizes the cooperative transmission for a single hop, i.e. within the defined network. Fig.~\ref{fig:ancol-video-topo} presents a representative network that is used for explaining several aspects of this work. Here we introduce the term \textit{communication phase} as the basic time frame for the protocol analysis in this paper. The proposed PHY works as follows. During the first part of this communication phase a number of $N$ network nodes broadcast their packets independently and concurrently. Due to the broadcast nature of the wireless channel the concurrently transmitted packets will be available to every node that does not transmit but it overhears the channel. In the second part of the communication phase the broadcasted and interfered packets are amplified and forwarded at the PHY by $N-1$ relays sequentially. The relay transmission order can be random but it should be decided in advance of a communication phase. It is also possible that one relay transmits the received signal $N-1$ times providing thus time diversity and reducing the number of required relays. The power constraint for the relays is the same with the original senders and this parameter is taken into account into our analysis. 

Regarding the lower layer aspects of our system, we assume full overlap between the transmitted packets. This is possible since transmissions use time division multiple access (TDMA) with pre-defined slot boundaries. At the receivers the packets at the PHY are decoded, after the final relay forwards the respective interfered packet, with a maximum-likelihood (ML) decoder that we describe in the next section. The optimality of interfering and decoding two packets for maximum throughput was studied and established in~\cite{katabi07a,argyriou:twc-ancol}. Therefore, the result of these works was that the cooperative system should allow the interference of two and not more packets for achieving throughput optimality.

Regarding the video streaming operations we consider the transmission of pre-compressed video where each video unit corresponds to a single I, P, or B frame~\cite{mihaela-book}. The first task of the video transmission system is to pass the video units through an application-layer FEC encoder in order to create the packet to be actually transmitted. In the proposed system each sender follows independently the previous process and broadcasts its respective packets. 

\section{Physical Layer with Over-the-Air Mixing}
\label{section:coop-phy}
\begin{figure}[t]
\begin{center}
\includegraphics[keepaspectratio,width = 1.0\linewidth]{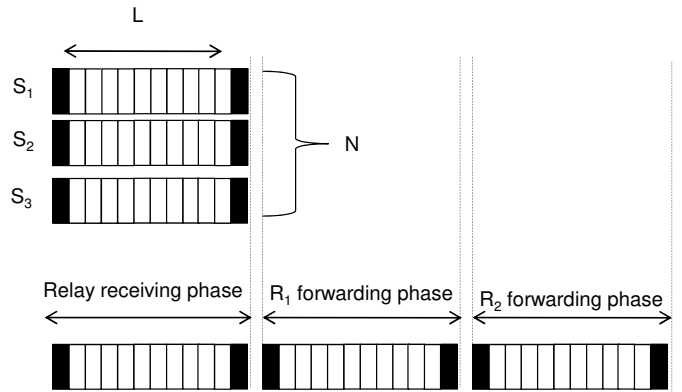}
  \caption{The AFOST cooperative protocol for the simple topology with three senders. The dark-shaded blocks indicate the symbols that belong to the preambles and postambles of a packet and are used for channel estimation, while the non-shaded blocks are the symbols of the information packet.}
  \label{fig:cooperative-protocol-af-st}
\end{center}
\end{figure}

For this PHY we present an example based on the topology depicted in Fig.~\ref{fig:ancol-video-topo}. Fig.~\ref{fig:cooperative-protocol-af-st} presents the corresponding protocol behavior in the time domain. In this example there is one broadcast phase from all the $N=$ 3 senders and two forwarding phases from all the $M=$ 2 relays. Because of spatial diversity, different versions of the broadcasted signals are received at different network nodes including the relays. During the forwarding phase, each participating relay broadcasts the locally received interfered signals after it applies the appropriate power scaling. An interesting observation is that the minimum number of required relaying phases is equal to the number of senders that transmit concurrently minus one. Therefore, it must be $M \geq N-1$. The reason is that if a number of nodes transmit concurrently, there is a need for at least the same number of forwarding phases so that $N$ linear equations are collected and the PHY decoder can then solve for the $N$ unknown and concurrently transmitted symbols. This communication scheme is named \emph{amplify and forward of over-the-air superimposed transmissions (AFOST)}.

\subsection{Description of AFOST}
%During the broadcast phase the senders are allowed to transmit by being synchronized only at the PHY symbol level.
Let $x_{n}$ denote the signal that is transmitted from the sender $n$. The number of PHY symbols $L_n$ denotes all the symbols of the packet that were transmitted during the broadcast phase. By using this notation, let us proceed with the analysis of the transmitted and received signals for the protocol we described. Recall that the group of senders that transmit concurrently is  $\mathcal{N}$. We may write the received signals at a relay $R_m$ as
\begin{eqnarray} \label{y_Rm}
y_{R_m}=\sum_{n \in \mathcal{N}} \sqrt{P}h_{S_n,R_m} x_{n} + w_{R_m}, \quad \forall m \in \mathcal{M},
\end{eqnarray}
where $\sqrt{P}$ is the transmission power at each sender, $h_{S_n,R_m}$ is channel transfer function between the sender and the relay, and $w_{R_m}\sim\mathcal{CN}(0,\sigma)$ denotes the AWGN at the relay $R_m$.
Similarly, the received signals at destination $D_k$ during the broadcast phase can be written as
\begin{eqnarray} \label{rpda1}
y_{D_k} & = & \sum_{n\in\mathcal{N}} \sqrt{P}h_{S_n,D_k} x_{n} + w_{D_k}.
\end{eqnarray}
For the one broadcast phase there are multiple forwarding phases, i.e. their precise number is $N-1$. In each of the forwarding phases a relay $R_m$ broadcasts the received signals given in~\eqref{y_Rm} by applying a power amplification factor $g_m$ so as to maintain the power constraint~\cite{laneman04}. The power gains are given as
\begin{eqnarray}\label{eqn:power-constraint1}
g_m =  \sqrt{\frac{P}{P\sum^{N}_{n=1}\gamma_{S_n,R_m} +\sigma^2}},
\end{eqnarray}
where $\gamma_i=|h_i|^2$. Subsequently, the relays forward the amplified signals. In the forwarding phase from $R_m$, the received signal at $D_k$ can now be written as
\begin{eqnarray}\label{eqn:y_Dk_Rm}
y_{D_k,R_m} &=& g_m h_{R_m,D_k}\sum_{n\in\mathcal{N}} \sqrt{P}h_{S_n,R_m} x_{n}+g_m h_{R_m,D_k} w_{R_m}+ w_{D_k}.
\label{fp1}
\end{eqnarray}
Note that~\eqref{eqn:y_Dk_Rm} corresponds to the received signal from one relay.

Based on the above analysis, we write in vector form the received signal for destination $D_k$ as follows
\begin{equation}
\mathbf{y}_{D_k} =\sqrt{P}\cdot\mathbf{G}\cdot\mathbf{H}_{D_k}\cdot\mathbf{x} +\mathbf{w}_{D_k}.
\end{equation}
The $N\times N$ channel matrix for the cooperative system we introduced is:
\begin{eqnarray}
\mathbf{H}_{D_k} =\left [ \begin{array}{ccc}
  h_{S_1,D_k} & .. &h_{S_N,D_{k}}\\
  h_{S_1,R_1}h_{R_1,D_k} &  ... & h_{S_N,R_1} h_{R_1,D_{k}}\\
  ... &  ... & ...\\
  h_{S_1,R_{N-1}} h_{R_{N-1},D_k} &  ... & h_{S_N,R_{N-1}} h_{R_{N-1},D_{k}}
  \end{array}\right] \nonumber
\end{eqnarray}
The array $\mathbf{G}$ corresponds to the power gains of all the relays and $\mathbf{w}_{D_k}$ is the noise vector that includes the broadcast and each forwarding phase (amplified noise). 

\subsection{PHY Decoding Algorithm}
\label{sec:ysic}
We now describe the PHY layer detection algorithm executed at each destination node. For the multi-user system we employ a minimum mean square error with successive interference cancellation (MMSE-SIC) receiver. If the Hermitian of $\mathbf{H}$ is $\mathbf{H}^H$, then the pseudo-inverse channel matrix $\mathbf{H}^{\dagger}=(\mathbf{H}^H \mathbf{H})^{-1}\mathbf{H}^H$ is used as follows: The MMSE approach tries to find a coefficient matrix $\mathbf{W}$ which minimizes the MMSE criterion. We have that $\mathbf{W}^{\dagger}=(\mathbf{H}^H \mathbf{H}+\sigma^2 \mathbf{I})^{-1}\mathbf{H}^H$. The $n-$th bit stream transmitted to destination node $D_k$ is extracted with the help of the pseudo-inverse channel matrix $\mathbf{W}^{\dagger}$ as follows. The signal is multiplied by the (estimated) pseudo-inverse:
\begin{eqnarray}\label{eq:y_tilde_mmse}
\hat{\mathbf{y}}_{D_k} &=&  \mathbf{W}^{\dagger}_{D_k,n} \mathbf{y}_{D_k}= \mathbf{W}^{\dagger}_{D_k,n} \mathbf{H}_{D_k,n}\mathbf{x}_k +\mathbf{W}^{\dagger}_{D_k,n}\mathbf{w}_k
\label{fp1}
\end{eqnarray}
where $\mathbf{W}_{D_k,n}^{\dagger}$ indicates \textit{all the rows in the pseudo-inverse channel matrix $\mathbf{W}_{D_k}^{\dagger}$ minus the $n$-th row}, while $\mathbf{H}_{D_k,n}$ symbolizes the $n-$th column for $\mathbf{H}_{D_k}$. Due to the whitening operation note that
$E[\mathbf{W}^{\dagger}_{D_k,n}\mathbf{w}_k|\mathbf{H}_{D_k,n}]=\Big ( \sum_{m=1}^{N-1}g_m^2|h_{R_m,D_k}|^2+1\Big )I_N$.
In addition, in our case we follow an MMSE-SIC where the power of the received signals in $\hat{\mathbf{y}}_{D_k}$ is ordered from higher to lower power. If we denote by $\mathbf{\bar{y}}$ the ordered version of the signal from higher to lower power of the received signals contained in $\mathbf{y}$ then we can apply the ordered SIC (OSIC) approach for detecting first the symbols that were received with the higher power. The destination uses MMSE equalization and estimates the higher power symbol (first in array $\mathbf{\bar{y}}$) $x_l$ as
\begin{eqnarray}\label{eq:x_hat_mmse}
 &&   \hat{\mathbf{x}}_{l,D_k} = \mathbf{W}_{D_k,1}^{\dagger} \mathbf{\bar{y}}_{D_k,1}
\end{eqnarray}
It is important to note that $\hat{\mathbf{x}}_{l,D_k}$ indicates the estimate of symbols from a sender $S_n$ but at node $D_k$. 

\subsection{Sum-Rate}
\label{section:sum-rate}
%\textbf{DO I NEED THESE: Pre-whitening along the lines described earlier, we obtain the whitened channel matrix  $\hat{\textbf{H}}_{A}^{(m+1\times n)}$ that it is}
%\begin{eqnarray}
%\left [ \begin{array}{ccc}
%  \sqrt{P}h_{1,d}/\sqrt{\sigma^2} & .. &\sqrt{P}h_{n,d}/\sqrt{\sigma^2}\\
%  \sqrt{P}h_{1,R_1}h_{R_1,d}/\sqrt{\lambda_1} &  ... & \sqrt{P}h_{n,R_1} h_{R_1,d}/\sqrt{\lambda_1}\\
%  ... &  ... & ...\\
%  \sqrt{P}h_{1,R_m} h_{R_m,d}/\sqrt{\lambda_m} &  ... & \sqrt{P}h_{n,R_m} h_{R_m,d}/\sqrt{\lambda_m}
%  \end{array}\right].
%\end{eqnarray}

For the MMSE/OSIC receiver that we adopted only the ergodic capacity or the average achievable rate can be practically calculated. Essentially it is the rate after averaging over a significant number of channel realizations. For the pair $S_n \rightarrow D_k$ we have that the average achievable rate will be:
\begin{equation}\label{eq:rate_estimate_mmse}
\widetilde{R}^{afost}_{D_k,S_n}= \mathbb{E} \Big[ \log_2 (1+\frac{P}{\sum_{m=1}^{N-1}g_m^2|h_{R_m,D_k}|^2+1}\mathbf{W}_{D_k,n}^{\dagger}\mathbf{H}_{D_k,n} )  \Big]
\end{equation}
%Note that in the above formula the expectation is not multiplied with the term $1/2$ since in $N$ transmission slots, $N$ packets are transmitted. 
This rate estimate is communicated from node $D_k$ to $S_n$ after it is estimated locally, since destination node $k$ has obtained the channel estimate $\mathbf{H}_{D_k,n}$.
%For all the channel realizations on all sender-destination pairs we have that if destination $D_k$ recovers bitstream $n$ then:
%\begin{equation}\label{eq:rate_estimate_mmse}
%\widetilde{R}^{mmse/osic}=\sum^{N}_{n=1,k=1} \widetilde{R}^{mmse/osic}_{D_k,S_n}
%\end{equation}

\section{Utility Optimization}
\label{section:utility-optimization}  In this section we attempt to optimize video transmission with a utility-based framework that uses the high-throughput cooperative protocol with superimposed packets that we presented in Section~\ref{section:coop-phy}.

\subsection{Utility Function}
We formulate our optimization problem as a utility maximization. Different utility functions can be employed by the senders. In our case, the utility function is defined as the reduction of the reconstruction distortion of the media presentation, i.e.,
\begin{eqnarray}\label{eqn:utility_def}
    u(r_i)=\sum_i \Delta D(i)\quad \text{with} \quad \sum_i \Delta R(i) \le T_{n,k},
\end{eqnarray}
where the RD information associated with packet $i$ consists of its size $\Delta R(i)$ in bytes and the importance of the packet for the overall reconstruction quality of the media presentation denoted as $\Delta D (i)$~\cite{ChakareskiF:06}. In practice, $\Delta D (i)$ is the total increase in the mean square error (MSE) distortion that will affect the video stream if the packet is not delivered to the client by its prescribed deadline. It is important to note at this point that the value of the MSE distortion in $\Delta D (i)$ includes both the distortion that is added when packet $i$ is lost and also the packets that have a decoding dependency with $i$. Now, in order to compute the utility $u(r_i)$ in (\ref{eqn:utility_def}) we previously label the media packets comprising the presentation in terms of importance using the procedure from \cite{ChakareskiF:06}. Therefore, the index $i$ in the summations in (\ref{eqn:utility_def}) enumerates the most important media packets in the presentation up to a data rate of $r_{i}$. In other words, $u(r_{i})$ corresponds to the cumulative utility of the most important packets up to the rate point $r_{i}$.

\subsection{Optimization for COOP/ORTH}
First we present the problem formulation for a PHY that uses cooperative transmissions (named COOP from now on) according to which each node transmits a packet with the help of an intermediate relay and in orthogonal time slots~\cite{laneman04}. The rate estimate for the cooperative PHY is a modified version of the Shannon capacity formula~\cite{laneman04}. For this COOP scheme we name the rate estimate $\widetilde{R}^{coop}_{n,k}$ and the effective throughput at the physical layer $T^{coop}_{n,k}$. The last estimate is used for the utility optimization step. This should be done such that the overall utility $U_{n,k}(j)$ of the GOP $j$ that belongs to the media flow that is transmitted over the wireless link $(n,k)$ is maximized.

Let us define as $c_{n,l}$ the TDMA slot allocation vector that indicates that the $n$-th user transmits in the $l$-th slot out of the $N$ maximum. Then the utility optimization problem is defined as:
\begin{eqnarray}
\label{eqn:optimization_problem_coop}
\max U_{n,k}(j) \quad \text{s.t.} \quad \left \{ \begin{array}{c}
  r_{n,k} \leq max (T^{dir}_{n,k},T^{coop}_{n,k})\\
  \sum^{N}_{n=1} \sum^{N}_{l=1} c_{n,l} = N \\
  c_{n,l}\in \{0,1\}\\
  U_{n,k} \in u(r_{i})
  \end{array}\right.
\end{eqnarray}
In the above, in the first constraint the best out of the direct or cooperative transmission modes is selected based on the rate estimate. The second and third constraints ensure that all the allocated slots to the $N$ transmitting nodes is equal to their number. The last constraint means that the maximized utility should consist of a valid R-D point that includes media packets up to packet $i$. This is necessary since there is a finite number of available rate points. The optimal solution to the above problem is out of the scope of this paper. Naturally, due to the complexity of the problem we resort here to a heuristic solution that does not require a centralized controller. More specifically, we allow the nodes to share equally the slots after every communication phase, and then they select locally the optimal transmission mode which is either direct (DIR) or COOP.

\subsection{Optimization for AFOST}
We use the formulation of the optimization problem given in~\eqref{eqn:optimization_problem_coop} and we adapt it given that the $AFOST$ protocol is now used. First of all since $AFOST$ is used during every transmission slot, only the residual rate constraint $T^{afost}_{n,k}$ is used. Second, every node transmits in each of the $N$ TDMA slots in a complete \emph{communication phase}. Therefore, the second/third conditions can be eliminated from~\eqref{eqn:optimization_problem_coop}. Using the notation introduced previously we can write the simpler optimization problem as
%\begin{eqnarray}
%\label{eqn:optimization_problem}
%& \max U_{n,k}(j), &  \\
%\text{s.t.} &  r_{n,k} \leq T_{n,k},\sum_{N} P_{n,k} \leq P_{tot}~\text{and}~U_{n,k} \in u(r_{i,j}). \nonumber
%\end{eqnarray}
\begin{eqnarray}
\label{eqn:optimization_problem}
\max U_{n,k}(j) \quad \text{s.t.} \quad \left \{ \begin{array}{c}
  r_{n,k} \leq T^{afost}_{n,k}\\
%  \sum_{N} P_{n,k} \leq P_{tot} \\
  U_{n,k} \in u(r_{i})
  \end{array}\right.
\end{eqnarray}
%The second constraint simply ensures that the transmit power of the nodes is within the total power budget of the system $P_{tot}$. \textbf{A critical observation here is that the power allocation constraint is linear since every node transmits at every time slot, contrary to e.g. OFDM systems where the use of sub-carriers introduces non-linearities~\cite{wong04}.}
It is evident here the simpler formulation of our problem. The decision to couple the proposed PHY with the utility optimization pays off at three levels: First, it enables a simpler solution algorithm will limited constraints, second it requires no central coordination for the slot allocation, and third it removes the non-linear constraint that dictates the maximum achieved rate based on the used PHY.

We proceed here by solving the optimization problem in \eqref{eqn:optimization_problem}. For this problem, we can apply Lagrange duality \cite{BoydV:04}
to the first constraint in (\ref{eqn:optimization_problem}) to produce the following partial Lagrangian
\begin{eqnarray}
\label{eqn:lagrangian1}
L_n(\lambda_n,r_{n,k}) & = & U_{n,k} - \lambda_{n,k} \cdot (r_{n,k} - T_{n,k}),
\end{eqnarray}
where $\lambda_{n,k} > 0$ is the Lagrange multiplier for link $n \rightarrow k$. Similarly, $r_{n,k}$ is current instantaneous rate allocation. Finally, $L_n(\lambda_{n},r_{n,k})$ represents the individual Lagrangian.
%We explain how we address the second constraint in (\ref{eqn:optimization_problem}) at the end of this section.
Now, (\ref{eqn:optimization_problem}) represents a concave optimization problem with linear constraints for the rate region as provided by
the link rate constraint in (\ref{eqn:optimization_problem}).
%%Therefore, strong duality holds with no
%%duality gap between the solutions for the primal and the dual optimization problems \cite{decomposition-tutorial}. We define the dual function as
%%\begin{equation}
%%g(\lambda_n)=\max_{r_{n,k}>0} L_n(\lambda_n,r_{n,k}), \label{eqn:dual_function}
%%\end{equation}
%%Hence, the dual optimization problem of (\ref{eqn:optimization_problem}) can be written as
%%%
%%\begin{eqnarray}
%%\label{eqn:dual_problem}
%%     && \min_{\lambda_n} g(\lambda_n), \\
%%     && \text{s.t.}~\lambda_{n,k} > 0. \nonumber
%%\end{eqnarray}
If $\lambda^*_n$ is the optimal solution for the dual problem, then the corresponding $r^*(\lambda^*_{n,k})$ is the solution to the primal problem defined in~(\ref{eqn:optimization_problem}).

It can be shown that the following two equations represent a solution for the primal-dual optimization problems. First, node $n$ computes the optimal rate allocation on link $n \rightarrow k$ using
\begin{eqnarray}%
    r^*_{n,k} = \arg \max_{r_{n,k}}\Big \{U_{n,k} - \lambda_{n,k} r_{n,k} \Big \}.
\label{eqn:Rs_opt}
\end{eqnarray} %
Then, given $r^*_{n,k}$ we employ a sub-gradient method to update the value of $\lambda_{n,k}$ as follows
\begin{equation} %
\lambda_{n,k} = \max\big\{0, \lambda_{n,k} +
\delta \Big(r^*_{n,k} - T_{n,k} \Big) \big\}.
\label{lagrange_multiplier}
\end{equation} %
In the above equation $\delta$ is a small constant. Sub-gradient adaptation methods such as (\ref{lagrange_multiplier}) are typically used in optimization problems involving Lagrange relaxation. %Lastly,~\eqref{eqn:Rs_opt} and~\eqref{lagrange_multiplier} are consecutively applied every time node $n$ performs rate allocation on its outgoing link. 

Now, as shown in \cite{ChakareskiF:06} (\ref{eqn:optimization_problem}) can be efficiently
solved using the rate-distortion characterization of the media packets comprising a flow. In particular, if $\Delta D(i_n)/\Delta R(i_n)$ is the utility gradient of packet $i_n$, i.e., packet $i$ from
flow $n$ then this packet is transmitted when $\Delta D(i_n)/\Delta R(i_n) > \lambda_{n,k}$. The above allows the transmission over the wireless channel only the most important packets such that the overall utility of the media flows is maximized.

%\begin{figure}[t]
%\begin{center}
%\subfigure[FOREMAN]{\includegraphics[keepaspectratio,width=0.8\linewidth]{gradient1_foreman.eps}}\hspace{-0.3cm}
%\subfigure[MOTHER \& DAUGHTER]{\includegraphics[keepaspectratio,width=0.8\linewidth]{gradient1_mother.eps}}
%  \caption{"Utility-per-bit" gradients for the two sequences under test.}
%  \label{fig:packet-gradient}
%\end{center}
%\end{figure}

\begin{figure}[t]
\begin{center}
  \subfigure[]{\includegraphics[keepaspectratio,width=0.5\linewidth]{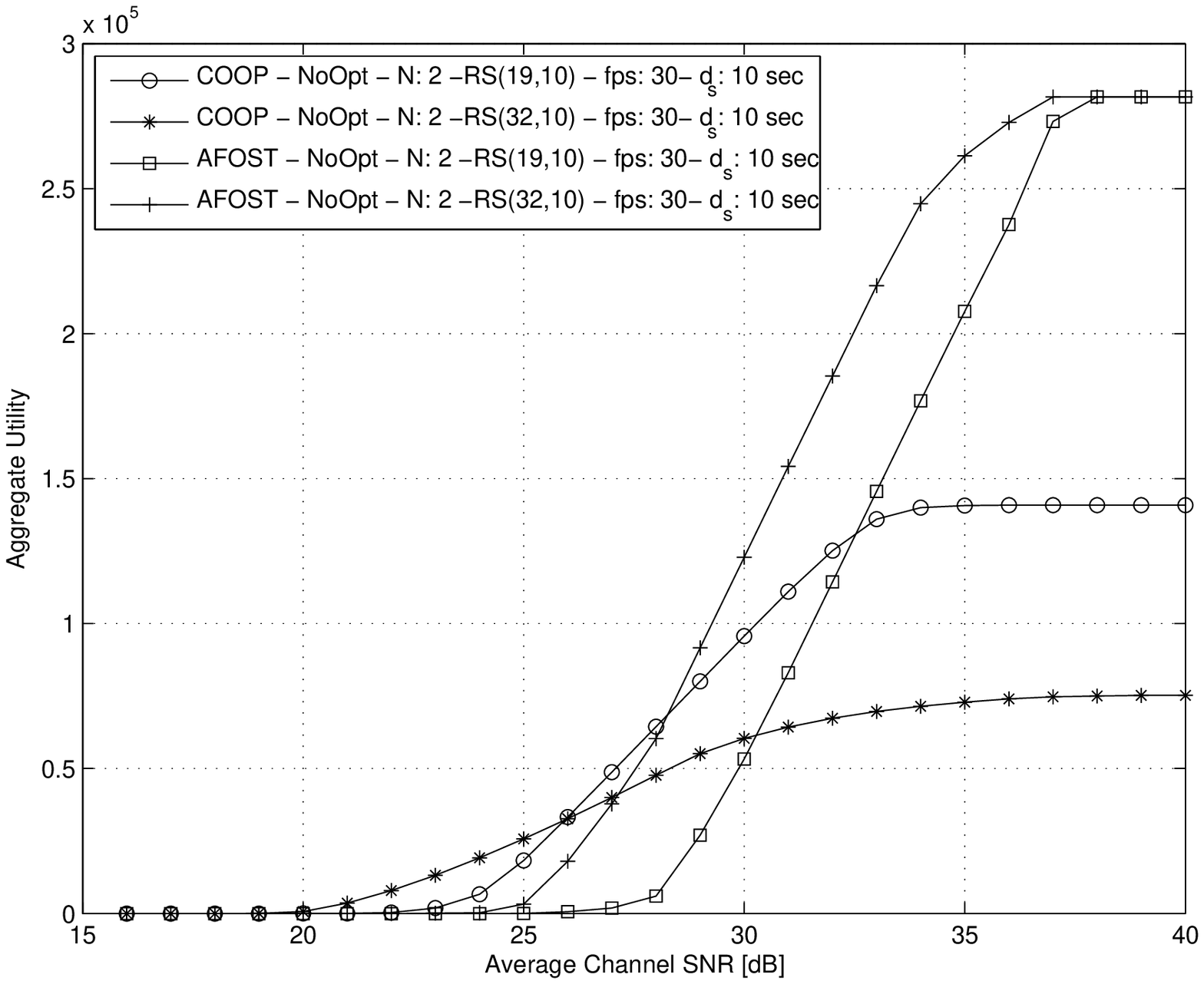}}\hspace{-0.0cm}%
  \subfigure[]{\includegraphics[keepaspectratio,width=0.5\linewidth]{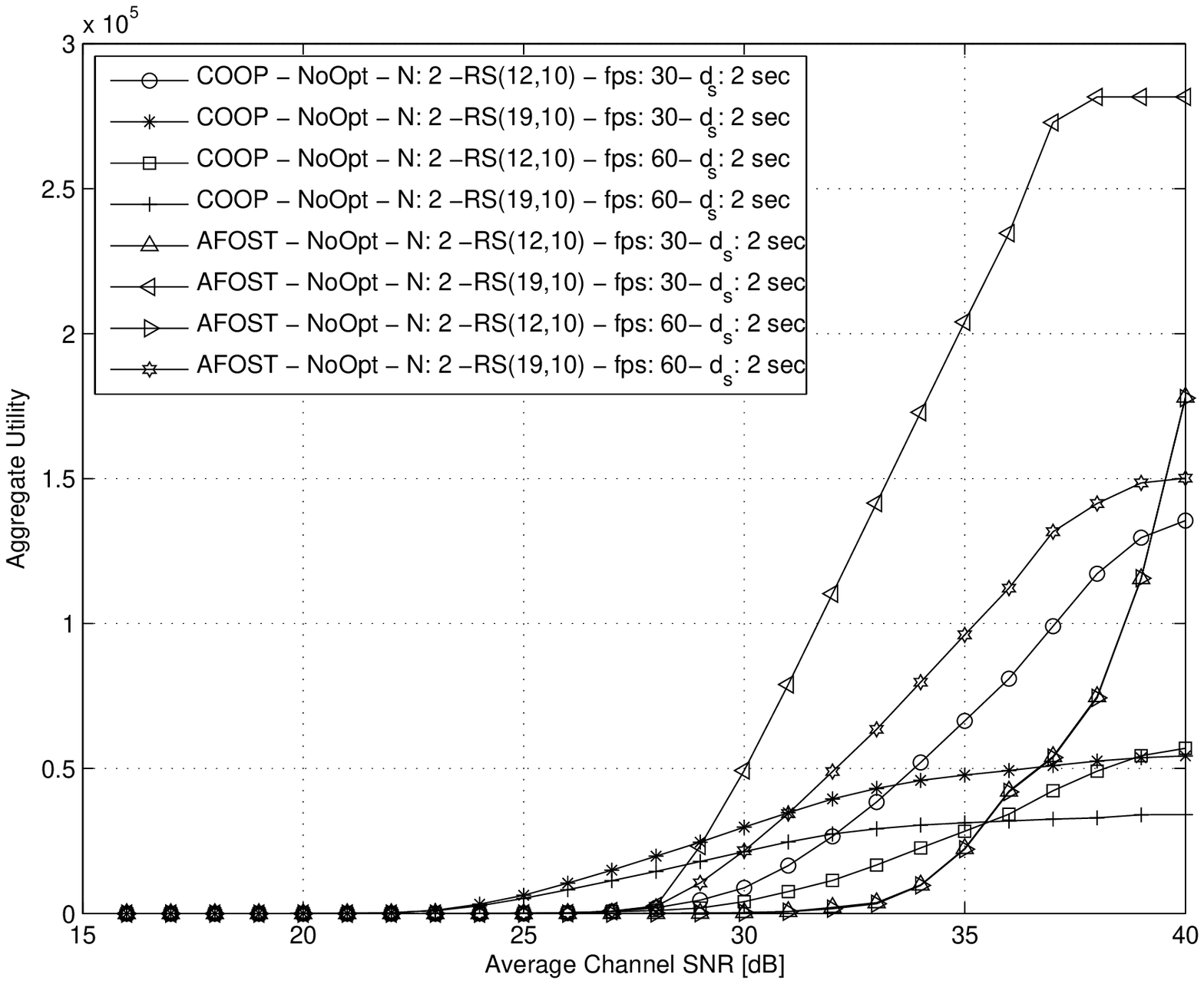}}%
    \caption{Average utility vs. the channel SNR.}
     \label{fig:results-no-opt}
\end{center}
\end{figure}

\section{Performance Evaluation}
\label{section:performance-evaluation} In this section, we present a comprehensive evaluation of the proposed system. We have implemented both the PHY and video streaming algorithms in Matlab. The number of nodes tested is kept small since the simulator operates at the PHY symbol level (not packet-level) requiring thus significant amount of running time. At the communication layer we tested a typical cooperative system that employs cooperative orthogonal amplify and forward at the PHY without collisions and it is named $COOP$ since the senders orthogonalize their transmissions~\cite{laneman04}. The proposed PHY with cooperative packet superimposition is named $AFOST$ in the figures. Regarding the lower layer parameters we assume a channel bandwidth of $W=20$ MHz, while the same Rayleigh fading path loss model was used for all the channels. The wireless link us frequency-flat fading and it remains invariant per transmitted PHY frame, but varies between simulated frames. The noise over the wireless spectrum is AWGN with variance $10^{-9}$ at every node. For the video part of the simulation, we examine the performances of two systems for media streaming, namely $NoOpt$ and $Opt$. The utility optimization was exercised for the duration of 10 GOPs. The media content used in the experiments consists of the CIF sequence MOTHER \& DAUGHTER that was compressed using an H.264 codec at the constant rate of 203 Kbps. Also 300 frames of the sequence were encoded at a frame rate of 30 or 60 fps using the following frame-type pattern IBBBP..., i.e., there are three B frames between every two P frames. The GOP size was set to 32 frames. Also, the startup/playback delay $d_s$ of the media presentation at every node is set according to the experiment.

\subsection{Results without Utility Optimization}
First, we present simulation results for the system where no utility optimization is applied, two users concurrently transmit ($N$=2), while packets are transmitted according to the presentation order. Fig.~\ref{fig:results-no-opt}(a,b) presents the related results for the $COOP$ and $AFOST$ systems. For a relatively high playback delay of $d_s$=10 sec, we see that $AFOST$ performs considerably better than $COOP$. Different application layer FEC code rate is needed for each system in order to achieve the best performance while other tested code rates provide worse performance. In this case $AFOST$ compensates with the higher code rate of $RS(32,10)$ the slightly increased BER. If the playback delay is even higher then the performance of both systems would converge. For $d_s$=2 sec and frame rate of 30fps shown in Fig.~\ref{fig:results-no-opt}(b), the performance of the $COOP$ system with orthogonal cooperative transmissions is significantly inferior to $AFOST$. At the high SNR regime, where the BER is reduced for both systems, the only choice is to reduce the FEC rate but this is not enough since the data rate of the communication link must be high in order to compensate for the short $d_s$. Only the $RS(12,10)$ code can provide some improvement for $COOP$. Therefore, $AFOST$ can support more effectively this higher data rate and low-delay requirement of multiple media streaming users. Also note that when the frame rate is increased to 60fps, both systems suffer due to the increased need for bandwidth but $AFOST$ still performs considerably better.  Another interesting result is that for the poor channel conditions, the $COOP$ system is still able to provide some meaningful aggregate utility contrary to $AFOST$. The reason is that this system can reduce the BER and it can provide at least some goodput to the application.

\begin{figure}[t]
\begin{center}
  \subfigure[]{\includegraphics[keepaspectratio,width=0.8\linewidth]{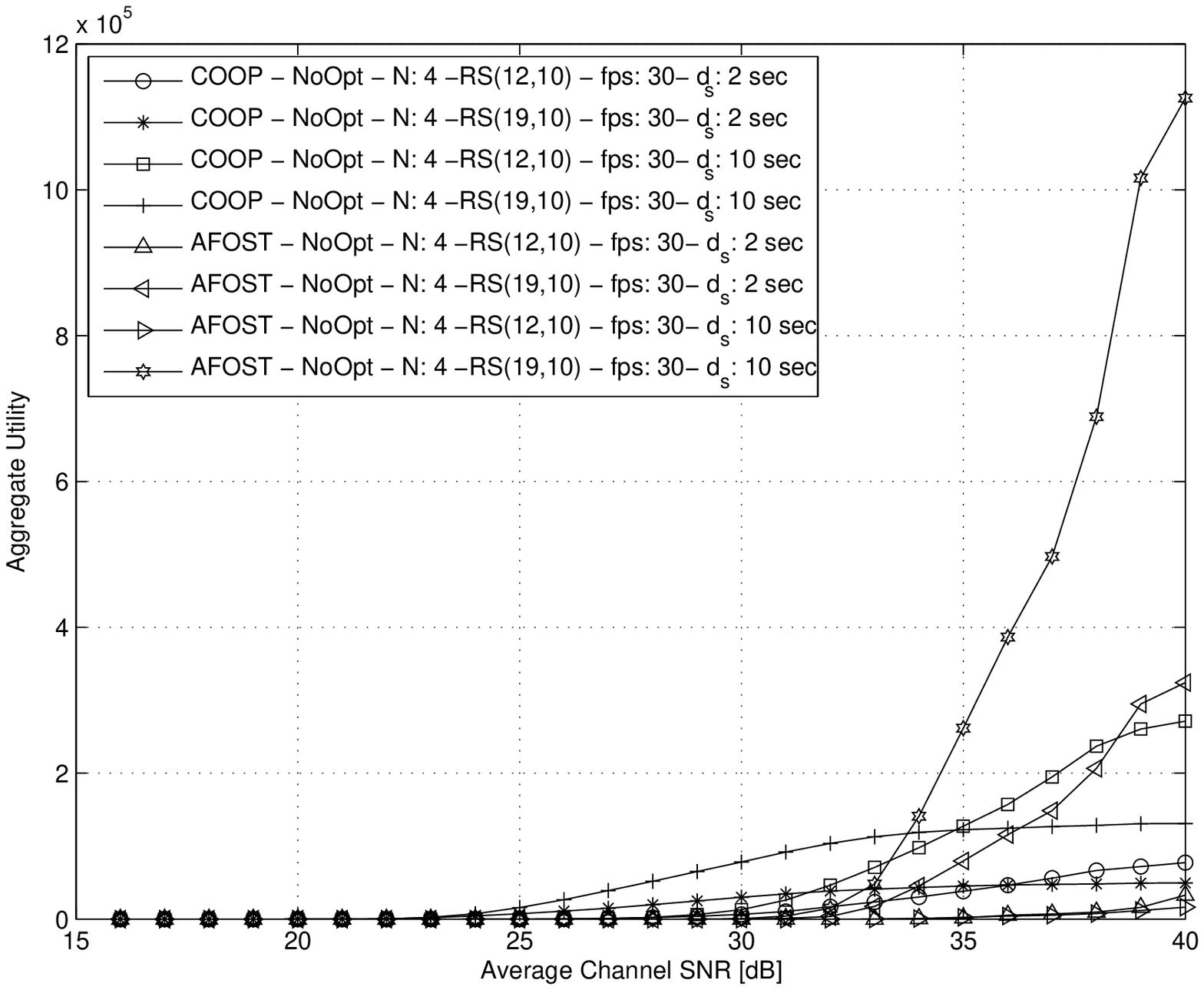}}%

%  \subfigure[]{\includegraphics[keepaspectratio,width=0.8\linewidth]{figures-matlab/test4_distortion.eps}}%

  \caption{Average utility vs. the channel SNR.}
  \label{fig:results-no-opt-four-nodes}
\end{center}
\end{figure}

The results for four nodes ($N$=4) can be seen in Fig.~\ref{fig:results-no-opt-four-nodes}. We see here that generally the $COOP$ system behaves considerably worse than $AFOST$ that achieves the best performance for a FEC code rate $RS(19,10)$. The situation is deteriorated for both systems when the startup delay $d_s$ is shorter by nearly an order of magnitude and equal to 2 sec. Still, the results are better for $AFOST$ over $COOP$.

\subsection{Results with Utility Optimization}
\begin{figure}[t]
\begin{center}
  \subfigure[]{\includegraphics[keepaspectratio,width=0.5\linewidth]{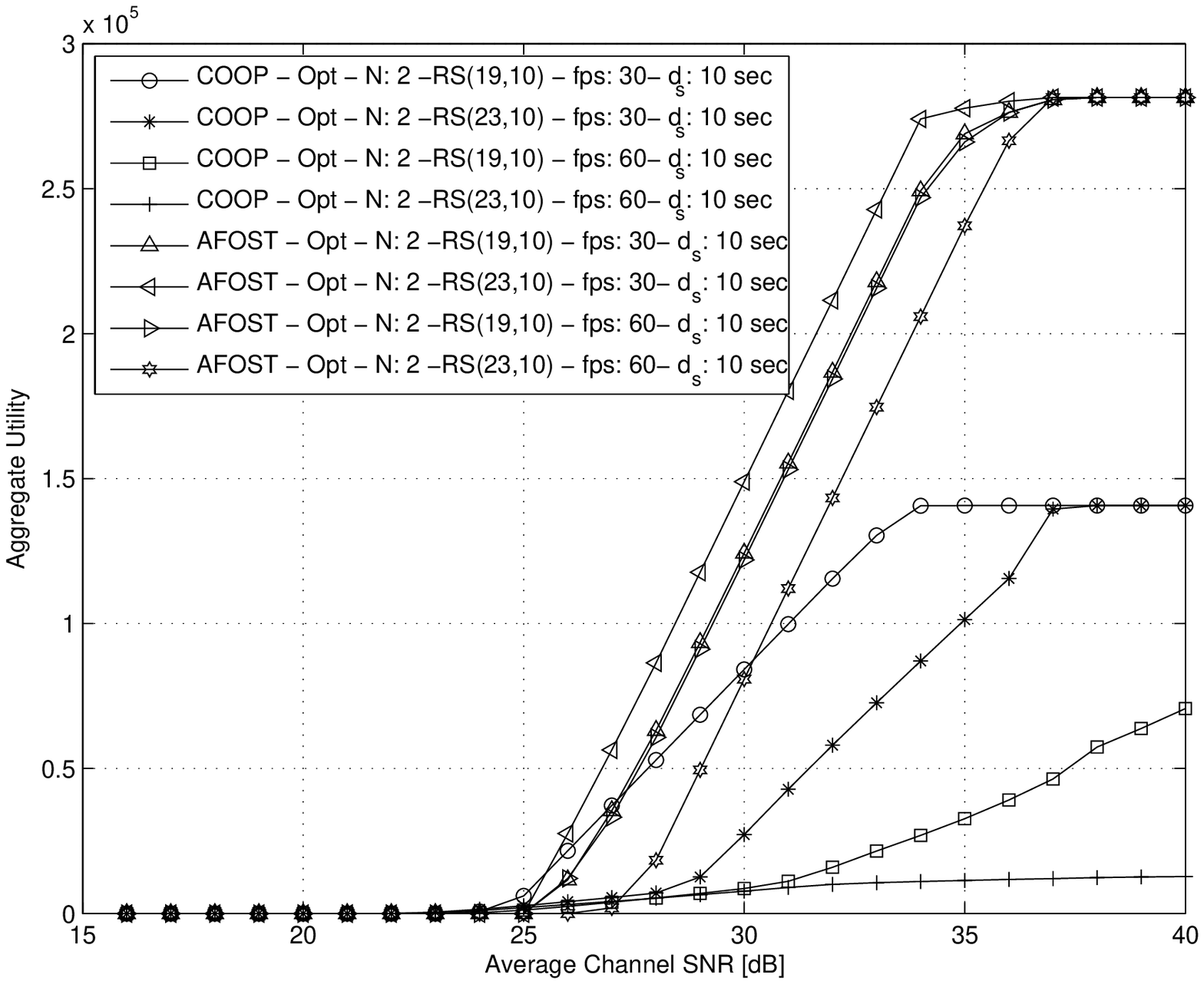}}\hspace{-0.0cm}%
  \subfigure[]{\includegraphics[keepaspectratio,width=0.5\linewidth]{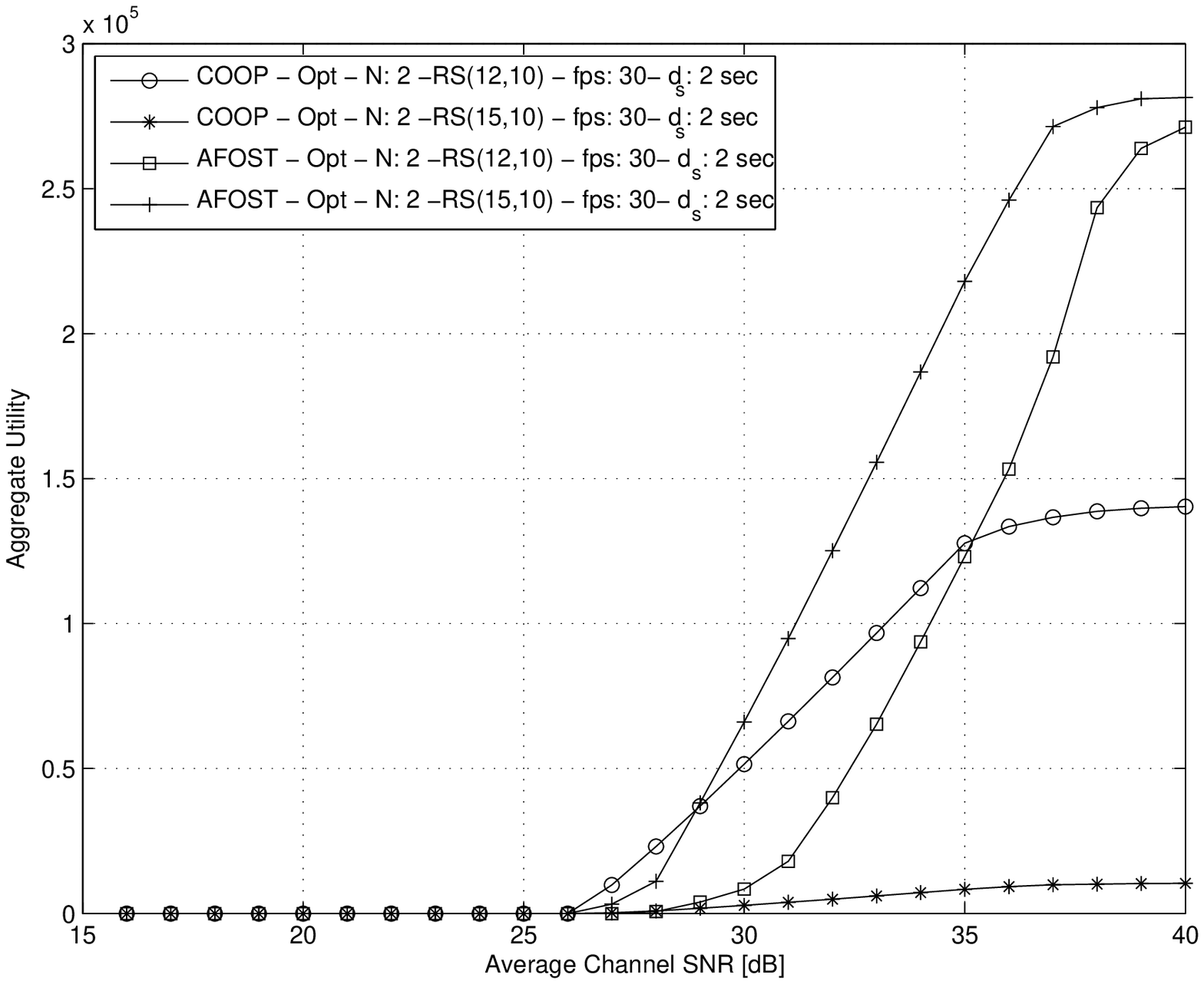}}%
   \caption{Average utility vs. the channel SNR.}
     \label{fig:results-opt}
\end{center}
\end{figure}

Fig.~\ref{fig:results-opt} presents results for the case that the utility optimization framework we developed in Section~\ref{section:utility-optimization} was enabled. Again, two concurrent senders are tested first for the $AFOST$ system. With the proposed utility optimization framework, the performance of the $COOP$ transmission mode is good only for the case of a more loose startup delay requirement of $d_s$=10 and 30 fps as Fig.~\ref{fig:results-opt}(a) indicates. $AFOST$ is superior on all tested cases. In the case of $d_s$=2, the $COOP$ system with $RS(12,10)$ is better than $AFOST$ for the same code rate, while the other FEC coding rate options under-perform significantly. But for a different FEC rate the proposed PHY again outperforms the orthogonal PHY. From these first results, and even for two users, two conclusions can already be made. First is that allocating the transmission rate and prioritizing important video packets improves considerably the video quality for $AFOST$, and second that utility optimization is crucial even if concurrent interfering transmissions with $AFOST$ are not enabled and a standard COOP is used.

\begin{figure}[t]
\begin{center}
  \subfigure[]{\includegraphics[keepaspectratio,width=0.5\linewidth]{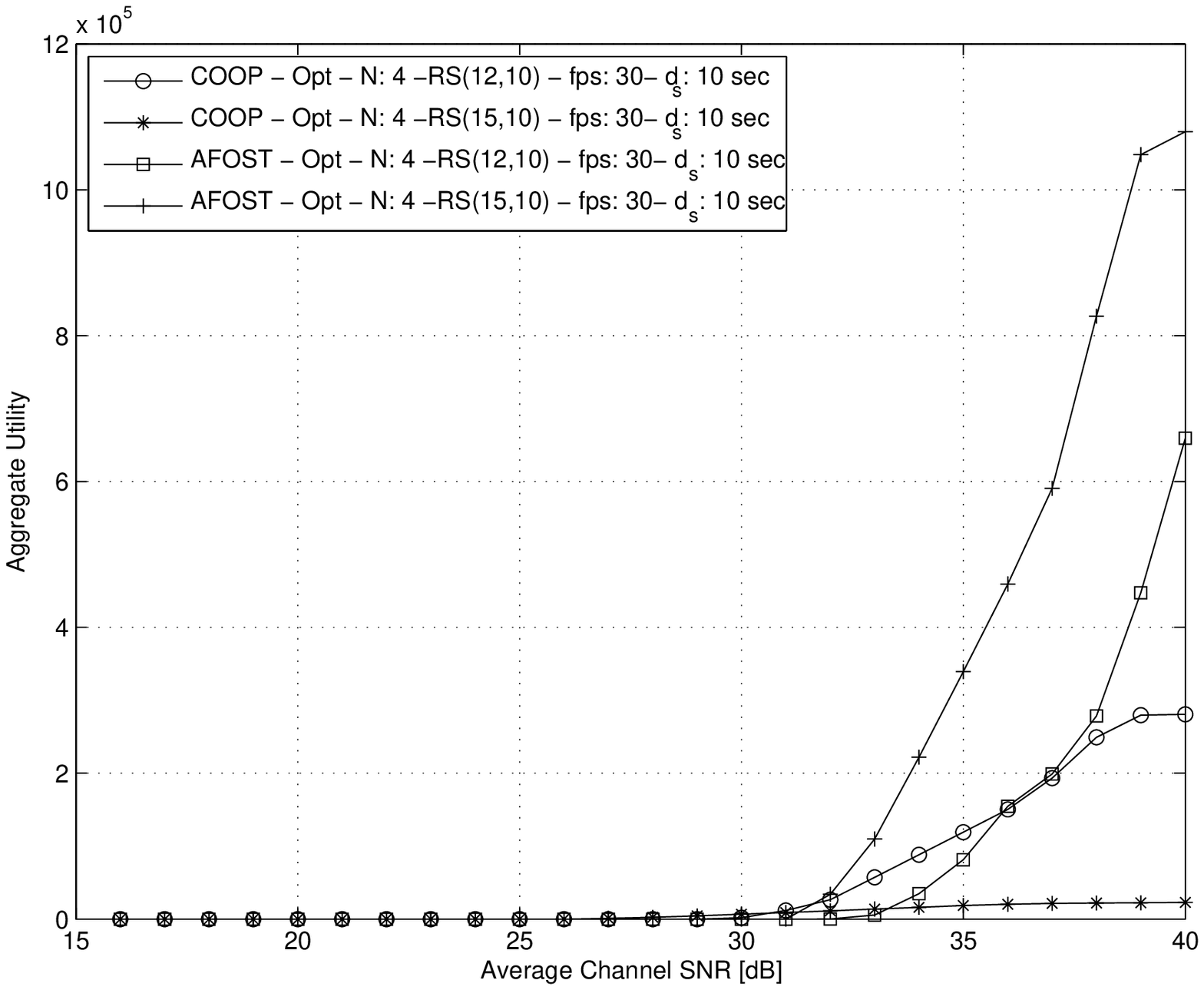}}\hspace{-0.0cm}%
  \subfigure[]{\includegraphics[keepaspectratio,width=0.5\linewidth]{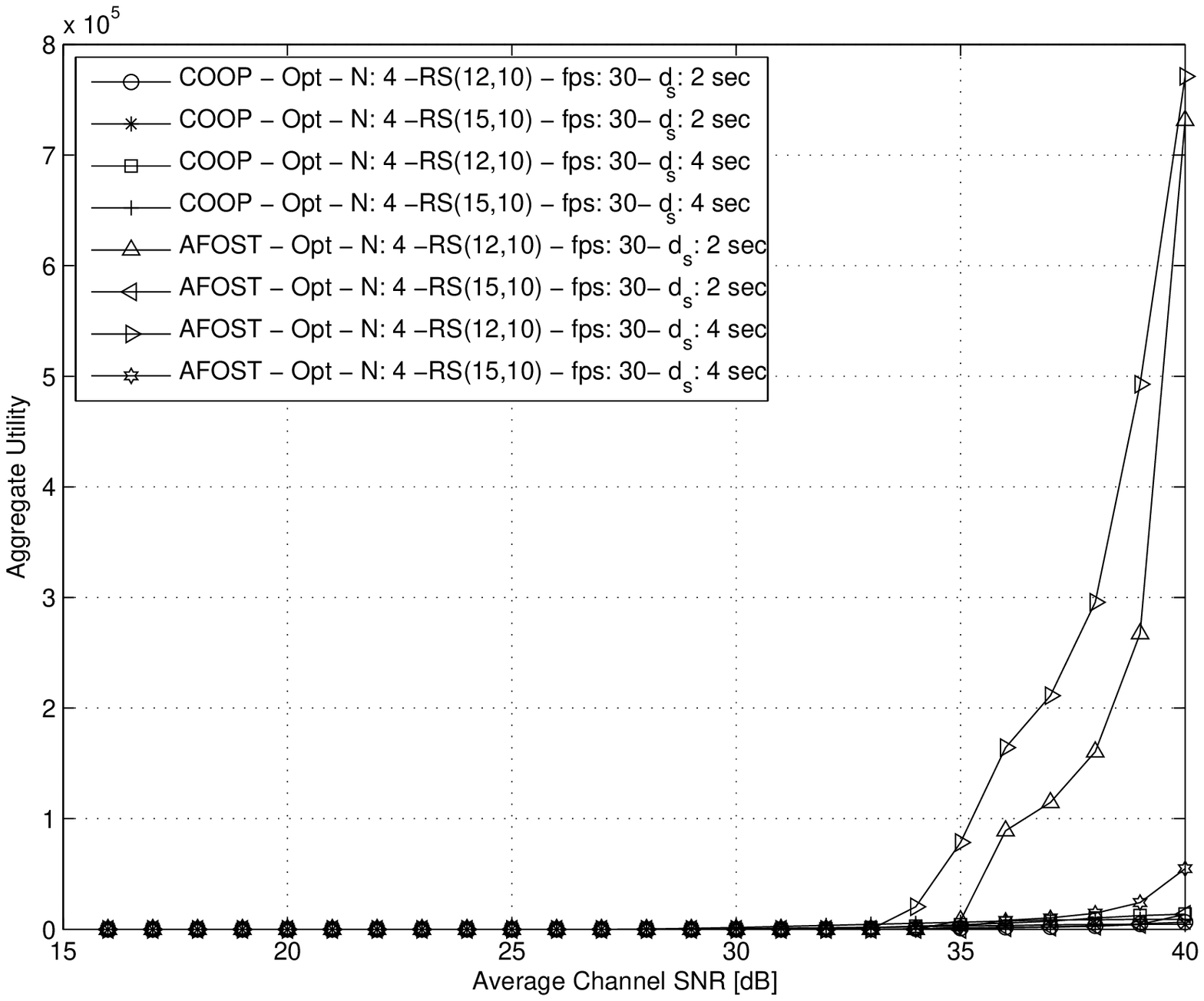}}%
  \caption{Average utility vs. the channel SNR.}
  \label{fig:results-opt-four-nodes}
\end{center}
\end{figure}

Very interesting results are obtained for $N$=4 and are shown in Fig.~\ref{fig:results-opt-four-nodes}(a,b). Although, $COOP$ performs relatively good (but still inferior to $AFOST$) for a high $d_s=10$, for $d_s$=2 it is nearly impossible to compete. The reason is that the required transmission rate is very high for transmitting the most important and high utility packets on-time. The most important media units are larger in number of bits (I and P frames) and so they require more bandwidth. So achieving higher throughput at the PHY of the communication stack is  more critical as more nodes share the medium. At the same time the aggregate utility is reaching significantly higher absolute values of nearly $8\times 10^5$ for good channel SNR. This behavior is attributed to the fact that the media units of highest importance are sent for all the four interfering senders. Furthermore, when we compare these results with the $NoOpt$ case, we see that $Opt$ outperforms $NoOpt$ for the case of four interfering nodes. Therefore, the option to allow collisions to occur between more than two nodes makes sense when video transmission is jointly employed with a RD utility optimization framework. Alternatively, when the $COOP$ transmission mode is selected, it is not so critical to employ $Opt$.

\section{Conclusions}
\label{section:conclusions}
In this paper we presented a wireless video transmission system that allows interfering transmissions to occur as part of the normal system operation. The first direct benefit was that when multiple senders transmit concurrently, throughput is improved since a higher number of transmitted packets per unit of time can be recovered at the PHY especially as the channel SNR improves. The second indirect benefit from the adoption of the new PHY was a simpler formulation of the utility optimization problem. The performance results showed that even though superimposing only two wireless packets is optimal for increasing throughput, the utility of multiple received video sequences can also be increased even if more than two senders transmit concurrently. The later result was also shown to be possible even if the utility optimization framework is disabled and only the proposed scheme is used. If a delay constraint is present, the previous result is emphasized even more since the concurrent transmission naturally expedites the transmission of a higher number of packets on-time. 

\bibliography{../tony-bib}

\end{document}